\documentclass[10pt,conference]{IEEEtran}
\IEEEoverridecommandlockouts
\usepackage{cite}
\usepackage{amsmath,amssymb}
\usepackage{graphicx}
\usepackage{booktabs}
\usepackage{etoolbox}
\patchcmd{\abstract}{---}{.\ }{}{\PackageError{pqc-corpus}{Abstract punctuation patch failed}{}}
\patchcmd{\IEEEkeywords}{---}{.\ }{}{\PackageError{pqc-corpus}{Keyword punctuation patch failed}{}}
\usepackage{array}
\usepackage[T1]{fontenc}
\usepackage{url}
\usepackage{xurl}
\usepackage{listings}
\usepackage{xcolor}
\usepackage[hidelinks]{hyperref}
\definecolor{patchremoved}{RGB}{145,35,40}
\definecolor{patchadded}{RGB}{20,95,55}
\hypersetup{pdftitle={PQC-Corpus: A Dataset of Linked Development Histories for Post-Quantum Cryptography},pdfauthor={Souhardya Saha Dip, Moshood A. Fakorede, Umar Farooq}}

\begin{document}
\bstctlcite{pqc_bstcontrol}
\title{PQC-Corpus: A Dataset of Linked Development Histories for Post-Quantum Cryptography}
\author{\IEEEauthorblockN{Souhardya Saha Dip\IEEEauthorrefmark{1}, Moshood A. Fakorede\IEEEauthorrefmark{1}, Umar Farooq}
\IEEEauthorblockA{Louisiana State University, Baton Rouge, Louisiana\\
\{ssahad1, mfakor1, ufarooq\}@lsu.edu}
\thanks{\IEEEauthorrefmark{1}These authors contributed equally to this work.}}
\maketitle

\begin{abstract}
As post-quantum cryptography (PQC) standards evolve, developers must adapt implementations, tests, and protocols. Studying this work requires connecting code changes to problem reports and developer reviews. We present PQC-Corpus, a dataset of linked development histories from 179 public GitHub repositories, spanning cryptographic libraries and general-purpose projects. It connects issues and closing pull requests to commits and full patches, with comments and reviews. We combine Git histories with GitHub threads, reconstruct 23 pull-request commit lists truncated at 250 entries, and check file integrity, schemas, and links. Two annotators independently assessed sampled commits and issues ($\kappa=0.86$ and $0.93$), labeling about seven in ten commits and two in three issues PQC-related within these samples. We provide full and filtered views with removal ledgers, and quantify how filtering reduces links from issues to commits. A patch-free view supports inspection before downloading patches, while traversal scripts follow issues to code changes. We make the dataset and collector publicly available for studies of specification changes, hybrid interoperability, cryptographic review, and migration assistance.

\end{abstract}

\section{Introduction}
Public-key cryptography supports secure communication and software signing, but a sufficiently powerful quantum computer could break RSA and elliptic-curve cryptography~\cite{shor1994algorithms}. Post-quantum cryptography (PQC) provides algorithms designed to resist classical and quantum attacks while running on conventional computers~\cite{chen2016reportpqc}. In August 2024, NIST finalized the ML-KEM, ML-DSA, and SLH-DSA standards~\cite{nist2024}. Industry efforts include Meta's hybrid TLS deployment~\cite{lin2024metatls}, AWS's service-migration plan~\cite{campagna2024awspqc}, Google's transition to ML-KEM in Chrome~\cite{adrian2024chrome}, and Apple's PQ3 protocol for iMessage~\cite{apple2024pq3}.

These integrations require developers to accommodate performance and resource costs~\cite{truong2026nistpqcstandardsreview} while preserving protocol interoperability~\cite{ott2019identifyingresearchchallengespost}. During Meta's rollout, larger Kyber keys added a round trip to some resumed TLS connections, and a \texttt{liboqs} race condition required an upstream fix~\cite{lin2024metatls}. N\"ather et al.'s review found sparse coverage of migration execution and maintenance~\cite{nather2024migrationreview}. Studying this work across projects requires linking reported problems to patches, tests, and reviews.

\textbf{Motivation.} Issue~\#46 in \texttt{RustCrypto/KEMs} requests alignment with final FIPS~203~\cite{rustcrypto2024issue46}. Its closing pull request (PR), \#47, includes a formal review and five commits that add domain separation to K-PKE key generation (Fig.~\ref{fig:patch}), clarify matrix-index ordering, adopt NIST's ACVP test vectors, and update draft-standard wording.

\begin{figure}[t]
{\footnotesize\ttfamily ml-kem/src/pke.rs\hfill ffd2239\par}
\lstset{basicstyle=\fontsize{8}{9}\selectfont\ttfamily,
  columns=fullflexible, keepspaces=true, showstringspaces=false,
  breaklines=true, breakatwhitespace=true, breakindent=1em,
  frame=tb, framerule=.3pt, rulecolor=\color{black!30},
  backgroundcolor=\color{black!3}, framesep=4pt,
  aboveskip=5pt, belowskip=0pt,
  xleftmargin=4pt, xrightmargin=4pt,
  framexleftmargin=4pt, framexrightmargin=4pt,
  morecomment=[l][\color{black!65}]{//},
  moredelim=**[is][\color{patchremoved}]{<removed>}{</removed>},
  moredelim=**[is][\color{patchadded}]{<added>}{</added>}}
\begin{lstlisting}
  pub fn generate(d: &B32)
      -> (Self, EncryptionKey<P>) {
    // Generate random seeds
<removed>-   let (rho, sigma) = G(&[d]);</removed>
<added>+   let k = P::K::U8;</added>
<added>+   let (rho, sigma) = G(&[&d[..], &[k]]);</added>

    // Sample pseudo-random matrix and vectors
    let A_hat: NttMatrix<P::K> =
        NttMatrix::sample_uniform(&rho, false);
\end{lstlisting}
\caption{Commit \texttt{ffd2239}, from PR~\#47 that closes issue~\#46, adds the parameter-set byte \texttt{k} for domain separation in K-PKE key generation. The excerpt is reindented and wrapped, with \texttt{-}/\texttt{+} marking removed and added lines.}
\vspace{-5pt}
\label{fig:patch}
\end{figure}

\begin{figure*}[t]
\centering
\includegraphics[width=\textwidth]{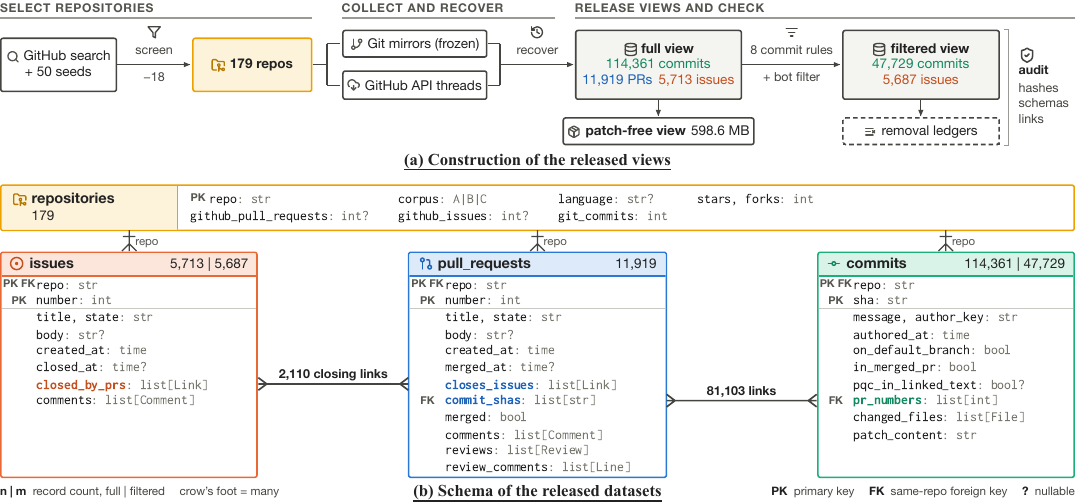}
\caption{Construction and JSONL schema of PQC-Corpus. Panel~(b) shows 41 of 71 top-level fields; \texttt{SCHEMA.md} defines all fields and nested objects.}
\label{fig:schema}
\end{figure*}

Existing PQC research examines readiness in source snapshots~\cite{mehl2026pqcreadiness} and Android apps~\cite{strauss2025}, organizational adoption~\cite{zimmermann2025organizationaladoptionpqc}, and quantum-software evolution~\cite{Upadhyay_2025}. PQC side-channel datasets provide power and electromagnetic traces~\cite{rezaeezade2025kyberpower,magazin2026openem,yuan2025dtds}. Migration benchmarks include 800 synthetic Python code pairs~\cite{pallares2026migration} and PQCMigrate-Bench's 10 tasks with held-out tests from open-source commit ranges~\cite{10.1145/3839338.3857266}.

GHTorrent and Software Heritage provide general repository data~\cite{gousios2013ghtorrent,pietri2019swh}, while PI-Link connects issues to PRs~\cite{alshara2023pilink} and vulnerability datasets link vulnerabilities to fixes or tests~\cite{bhandari2021cvefixes,nikitopoulos2021crossvul,vul4j2022}. SWE-bench~\cite{jimenez2024swebenchlanguagemodelsresolve} and MobileDev-Bench~\cite{fakorede2026mobiledevbenchbenchmarkissueresolution} pair issues with developer patches and tests to evaluate issue resolution.

\textbf{Contributions.} PQC-Corpus~\cite{pqccorpus_data} provides:
\begin{itemize}
\item \textbf{Linked development evidence} from 179 repositories, including patches, issues, PR comments, and reviews of merged and unmerged work.
\item \textbf{Documented collection and filtering}, with repository scopes, full and filtered views, removal ledgers, integrity checks, and PQC-relevance labels for sampled commits and issues.
\item \textbf{Resources for offline reuse}, including a patch-free view, traversal and characterization scripts, and open collection tools.
\end{itemize}

\section{Dataset Construction}
\subsection{Repository Selection}
We combined 50 curated seed repositories with 129 found through GitHub topic, metadata, and organization searches. Searches required PQC-related metadata and at least 30 stars, excluding forks and archived or disabled projects. One author screened candidates and excluded 18 for name collisions, non-software content, or absent PQC content. Seeds such as Go, OpenSSL, Kubernetes, and Bouncy Castle bypassed metadata screening to include PQC work in general-purpose projects.

A repository's score is the fraction of issues and PRs matching any of 90 PQC terms in their title, body, or comments. We calibrated the cutoff using 24 PQC-focused and 17 general-purpose anchor repositories. Any cutoff above the highest general-purpose score ($552/5880\approx0.094$) and up to the lowest focused score ($4/27\approx0.148$) separates the anchors. We chose the upper endpoint, rounded to 0.148, to limit unrelated records.

For repositories at or above the cutoff, corpus~A targets all issues, PRs, and reachable Git history. Below the cutoff, corpus~B retains issues and PRs matching a term in their \emph{title or body}, plus commits with matching messages or membership in a retained PR. Corpus~C contributes Git history from repositories with no issues or PRs. The release contains 81 A, 87 B, and 11 C repositories.

\subsection{Collection and Recovery}
The open pipeline~\cite{pqccorpus_code} mines frozen Git mirrors with PyDriller~\cite{10.1145/3236024.3264598} and uses a Deno~\cite{deno_runtime} REST/GraphQL collector to retrieve issues, PRs, comments, formal reviews, and inline review comments for retained threads. It partitions searches exceeding GitHub's 1,000-result cap and resumes from SQLite checkpoints. Pre-selection inventories contain 191,431 PRs, 171,911 issues, and 1,068,427 Git commits, with per-corpus counts in the supplement.

GitHub caps PR--commit lists at 250 entries, so the collector rebuilds longer lists from Git objects. It accepts a list only if its length matches GitHub's count and it contains every API-returned SHA. It fetches missing commits into a separate recovery mirror to preserve the frozen source and reports unresolved gaps. The release contains 23 rebuilt lists across six repositories, with 23,075 links beyond the cap (28.5\% of all PR--commit links). No released list has exactly 250 commits.

Acquisition timestamps span September 21--24, 2026. Filtering observations saved on September 29 record default-branch reachability, merged-PR membership, and whether linked text mentions PQC terms.

\begin{figure*}[t]
\centering
\begin{minipage}[t]{\columnwidth}
\vspace{0pt}
\centering
\includegraphics[width=\linewidth]{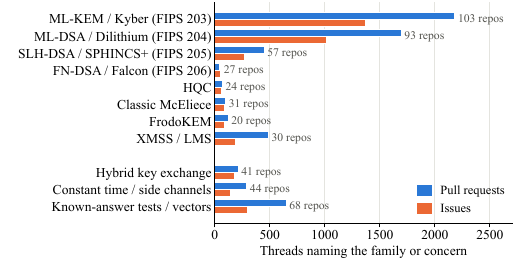}
\caption{PR and issue counts by algorithm family or engineering concern. Threads can match several categories, and annotations count repositories with a matching PR or issue.}
\vspace{-5pt}
\label{fig:coverage}
\end{minipage}\hfill
\begin{minipage}[t]{\columnwidth}
\vspace{0pt}
\centering
\includegraphics[width=\linewidth]{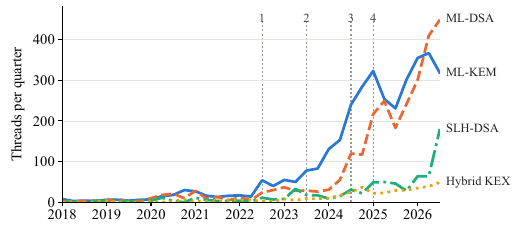}
\caption{Quarterly PR and issue counts matching three algorithm families and hybrid key exchange. Markers indicate (1)~NIST selection in Jul.\ 2022, (2)~draft FIPS 203--205 in Aug.\ 2023, (3)~final FIPS 203--205 in Aug.\ 2024, and (4)~HQC selection in Mar.\ 2025. Counts for 2026 Q3 end on September 24.}
\vspace{-5pt}
\label{fig:activity}
\end{minipage}
\end{figure*}

\subsection{Full and Filtered Views}
The full view retains all collected records, including unmerged work. Eight ordered rules produce a filtered view for studying integrated changes.

\noindent \textbf{Eligibility.} (1) We retain commits reachable from the default branch or associated with a merged PR.

\noindent \textbf{Repeated changes.} We remove (2) rebase copies, (3) merges repeating non-merge patches, (4) squash copies, and (5) merges whose second parent is already retained. Rules 2 and 4 apply only when patch and metadata comparisons identify exactly one matching commit reachable from the default branch.

\noindent \textbf{Background activity.} A separate PQC term matcher checks commit messages, linked PR/issue titles and bodies, paths, and changed lines. We remove non-matching commits (6) from GitHub forks, then (7) from repositories with no PQC terms in their names, descriptions, or topics.

\noindent \textbf{Housekeeping.} (8) We remove commits changing no files. When the diff lacks PQC terms, we also remove commits with bot author or committer names, or commits changing only CI files, only lock files, or only repository metadata.

Filtering retains 47,729 of 114,361 commits. We retain commits with ambiguous duplicate matches and skip patch-based keyword and identity checks for patches above 64~MiB. Separately, we remove 26 issues whose opener logins end in \texttt{[bot]}, leaving 5,687 issues. PRs and repository metadata remain unchanged.

\section{Data Model and Characterization}
\subsection{Storage and Joins}
Six UTF-8 JSON Lines (JSONL) views store repositories, PRs, and full and filtered commits and issues under four JSON Schemas (Fig.~\ref{fig:schema}b). The SQLite edition stores the same release in tables, separating nested comments, reviews, and links for SQL queries. Its filtered tables reference records in the full tables. Patches record first-parent changes, including binary changes, with base64 encoding for invalid UTF-8.

To trace an issue to its code changes in JSONL, find the PRs in its \texttt{closed\_by\_prs} list using each link's repository and number. Then use each PR's repository and \texttt{commit\_shas} to find its commits and patches. A PR can close an issue in another repository, so use the repository named in the link. PR records retain links even when the issue was not collected. The full view has 108 such links, which inner joins omit.

\subsection{PQC Coverage}
We search PR and issue titles and bodies case-insensitively for eight algorithm families and three engineering concerns after removing HTML comments and question-style checklist lines from bodies. Family patterns match 4,234 PRs (35.5\%) and 2,466 issues (43.2\%), including 2,181 PRs mentioning ML-KEM/Kyber across 85 repositories (Fig.~\ref{fig:coverage}). Hybrid key exchange, constant-time/side-channel terms, and test-vector terms appear in 41, 44, and 68 repositories, respectively. Fig.~\ref{fig:activity} plots thread creation against standardization milestones. Across all collected PRs, annual counts are 873, 1,947, 3,139, and 3,449 for 2023--2026, with 2026 counts through September 24. These counts describe text mentions and activity, without establishing implementation, cryptographic correctness, deployment, or causal effects of standardization.

\subsection{Repositories, Comments, and Reviews}
The dataset spans 29 GitHub primary-language labels, with commits from 167 repositories, PRs from 132, and issues from 128. Median pre-selection inventories were 1,466 commits and 176 PRs for corpus~B, compared with 178 commits and 18 PRs for~A. Corpus~C has a median of 14 commits and a median age of 0.7 years (supplementary plots). The five largest repositories by commit count in each view account for 51.0\% of full commits and 38.7\% of filtered commits. The 67,686 comments comprise 16,473 issue, 22,480 general PR, and 28,733 inline review comments. Across 7,302 PRs, there are 37,648 formal reviews. The \texttt{body} field is null or empty for 9.9\% of issues and 13.0\% of PRs.

\subsection{Choosing a View}
A complete issue--PR--commit path requires the chosen view to contain an issue, a PR that closes it, and one of that PR's commits. Filtering retains 99.5\% of issues, but only 58.9\% of those with a complete path in the full view retain one after filtering (Table~\ref{tab:linked-context}). Paths can share issues, PRs, and commits, so they are not independent examples.

\begin{table}[t]
\caption{Retention of linked development context after filtering}
\label{tab:linked-context}
\centering\footnotesize
\setlength{\tabcolsep}{3pt}
\begin{tabular}{@{}lrrr@{}}
\toprule
\textbf{Records, links, and paths} & Full & Filtered & Retained \\
\midrule
Issue records & 5,713 & 5,687 & 99.5\% \\
PR--commit links & 81,103 & 36,647 & 45.2\% \\
Complete issue--PR--commit paths & 10,081 & 6,183 & 61.3\% \\
\midrule
\multicolumn{4}{@{}l}{\textbf{Distinct records on complete paths}} \\
Issues & 1,865 & 1,099 & 58.9\% \\
PRs & 1,771 & 945 & 53.4\% \\
Commits & 7,637 & 4,547 & 59.5\% \\
\bottomrule
\end{tabular}
\end{table}

\subsection{Integrity Audit}
We checked file sizes, row counts, and SHA-256 hashes against the manifest, together with schema conformance, key uniqueness within repositories, and repository references. All 81,103 PR--commit links and 2,002 PR--issue closing links between collected records agree in both directions. Filtered records are byte-identical to their full-view counterparts. CSV ledgers record removal reasons for 66,632 commits and 26 issues, and filtering flags match the September 29 observations. These checks did not rerun filtering or assess historical completeness or cryptographic correctness.

\subsection{PQC Relevance}
\label{sec:relevance}
We sampled 100 full-view commits across PQC schemes and creation periods, with at most 20 per repository (50 from corpus~A, 44 from~B, and 6 from~C). The seeded issue sample contains 50 records each from A and B, spread across creation years with at most 15 per repository. Two authors independently applied the released codebooks~\cite{pqccorpus_code} to assess whether commits change, test, document, package, or discuss named PQC schemes or PQC-specific mechanisms, and whether issues concern PQC implementation, integration, or migration. They agreed on 94 commits and 97 issues (Cohen's $\kappa=0.86$ and $0.93$) and resolved nine disagreements through discussion. Table~\ref{tab:relevance} includes the 80 labeled commits retained after filtering. All 100 sampled issues remain in the filtered view. Neither sample is simple random, so these counts describe only the 200 labeled records and do not estimate corpus-wide relevance.

\begin{table}[t]
\caption{PQC-related records in the labeled samples}
\label{tab:relevance}
\centering\footnotesize
\setlength{\tabcolsep}{4pt}
\begin{tabular}{@{}lrrrrr@{}}
\toprule
\textbf{Sample} & $\kappa$ & \textbf{All} & \textbf{A} & \textbf{B} & \textbf{C} \\
\midrule
Commits, full view & 0.86 & 72/100 & 42/50 & 28/44 & 2/6 \\
Commits, filtered view & -- & 61/80 & 33/39 & 26/35 & 2/6 \\
Issues & 0.93 & 67/100 & 32/50 & 35/50 & -- \\
\bottomrule
\end{tabular}
\end{table}

\section{Dataset Usage}

\subsection{Tracing an Issue to Its Changes}
From the source supplement, reproduce the introductory example against the full JSONL release:
\begin{quote}\footnotesize\ttfamily
python3 scripts/reuse/trace\_issue.py \textbackslash\\
\hspace*{1em}-{}-data ../data \textbackslash\\
\hspace*{1em}-{}-repo RustCrypto/KEMs -{}-issue 46
\end{quote}
The script follows \texttt{closed\_by\_prs} to PR~\#47 and its five commits. The PR author marked it as work in progress while awaiting updated test vectors. After the issue's author linked NIST's new ACVP vectors, commit \texttt{6337c00} replaced the old vectors with ACVP JSON files. Another contributor then asked to keep the 831~KB and 558~KB files out of the published crate, and three minutes later commit \texttt{a96cf4b} excluded them and their tests. For issue~\#46, \texttt{-{}-filtered} returns the same commits and patches. Commit-message selection also retains Go commit \texttt{4b7f7cd87dfc}, ``crypto/tls: implement X25519MLKEM768,'' which links to a Gerrit review but has no collected PR.

\subsection{Prior Use}
We used PQC-Corpus in a study currently under review. The public artifact~\cite{pqccorpus_code} includes the relevance annotations and codebooks of Section~\ref{sec:relevance} and scripts that reproduce this paper's coverage and linked-context analyses.

\subsection{Research Enabled}
\noindent \textbf{Specification changes.} Trace how projects carry draft-to-final changes, such as Kyber to ML-KEM, into code, tests, and documentation using linked patches and dated threads around the milestones in Fig.~\ref{fig:activity}.

\noindent \textbf{Hybrid interoperability.} Compare integration and testing problems in the 392 matching threads across 41 repositories using proposals, reviews, and linked patches.

\noindent \textbf{Cryptographic review.} Examine how developers review constant-time behavior and test-vector changes using thread text, review states, and inline comments.

\noindent \textbf{Migration assistance.} Evaluate retrieval or generation of changes from issue text. Benchmark builders must validate target patches, hide linked PR and commit identifiers, exclude comments posted after the task cutoff, and split examples by repository and time.

\section{Limitations}
PQC-Corpus covers a purposive sample of public GitHub repositories. Corpus~B omits threads without matching terms, and all three corpora include records unrelated to PQC (Section~\ref{sec:relevance}), so researchers must screen records and linked changes for their task. Collection spans several dates, and later crawls reflect upstream changes.

\section{Conclusion}
In this paper, we introduce PQC-Corpus, a public dataset of post-quantum cryptography development histories from 179 GitHub repositories, spanning cryptographic libraries and general-purpose projects. It links issues and closing PRs to commits and full patches, with comments and reviews for studying how developers implement, test, and revise PQC integrations.

We provide full and filtered views, a patch-free view, removal ledgers, and relevance labels for sampled records, along with the collector and scripts for tracing issues to changes. Researchers can compare responses to standards changes, developers can examine implementation decisions, and tool builders can select and validate migration examples.

\textbf{Future work.} We intend to expand the corpus with additional repositories and successive snapshots to track responses to later standards changes. Task-specific relevance annotations would support sampling for migration benchmarks.

\section{Data Availability}
We provide PQC-Corpus as JSONL on Hugging Face (revision \texttt{4d72fb0})~\cite{pqccorpus_data} and a SQLite database on Zenodo~\cite{pqccorpus_zenodo}. The collector, patch-free ZIP, relevance annotations, and reuse scripts are available on GitHub~\cite{pqccorpus_code}. The collector and ZIP are also archived on Zenodo~\cite{pqccorpus_code_zenodo}.

\section*{Acknowledgments}
This work was supported by the National Science Foundation (NSF) under CNS-2540631. Opinions expressed are those of the authors and do not necessarily reflect the views of the NSF. We thank the maintainers and contributors of the projects represented in PQC-Corpus.

\clearpage
\bibliographystyle{IEEEtran}
\bibliography{references}
\end{document}